\documentclass[12pt,preprint]{aastex}

\def\a4{\hsize 17.0cm \vsize 25.cm}
\newcommand{\der}[2]  { \frac{{\rm d}#1}{{\rm d}#2} }

\newcommand{\dif}     {{\rm d}}

\shorttitle{2D Model of a Bimodal Wind}
\shortauthors{ W\"unsch et al.}

\begin{document}

\title{Supersonic Line Broadening within Young and Massive  Super Star 
Clusters}

\author{Guillermo Tenorio-Tagle\altaffilmark{1,2}, Richard 
W\"unsch\altaffilmark{3,4}, Sergiy Silich\altaffilmark{1}, 
Casiana Mu\~noz-Tu\~n\'on\altaffilmark{5,6}}
\and
\author{Jan Palou\v{s}\altaffilmark{4}}
\altaffiltext{1}{Instituto Nacional de Astrof\'\i sica Optica y
Electr\'onica, AP 51, 72000 Puebla, M\'exico; gtt@inaoep.mx}
\altaffiltext{2}{Sackler Visiting Fellow, Institute of Astronomy, University of Cambridge, U. K.}
\altaffiltext{3}{Cardiff University, Queens Buildings, The Parade, Cardiff. 
CF24 3AA; richard@wunsch.cz}
\altaffiltext{4}{Astronomical Institute, Academy of Sciences of the Czech
Republic, Bo\v{c}n\'\i\ II 1401, 141 31 Prague, Czech Republic}
\altaffiltext{5}{Instituto de Astrof\'{\i}sica de Canarias, E 38200 La
Laguna, Tenerife, Spain; cmt@ll.iac.es}
\altaffiltext{6}{Departamento de Astrofisica, Universidad de La Laguna, 
E-38205, La Laguna, Tenerife, Spain}

\begin{abstract}
The origin of supersonic infrared and radio recombination nebular lines 
often detected in  young and massive superstar clusters are discussed.
We suggest that these arise from a collection of repressurizing shocks 
(RSs), acting effectively to re-establish pressure balance within the  
cluster volume and from the cluster wind  which leads to an even broader although much weaker component. The supersonic lines are here shown 
to occur in clusters that  
undergo a bimodal hydrodynamic solution (Tenorio-Tagle et al. 2007), that is 
within clusters that  are 
above the threshold line in the mechanical luminosity or cluster  mass  vs 
the size of the cluster (Silich et al. 2004). The plethora of repressurizing 
shocks is due to  frequent and recurrent thermal instabilities that take 
place within the matter reinserted by stellar winds and supernovae. 
We show that the maximum speed of the RSs and of the cluster wind, are both 
functions of the temperature reached 
at the stagnation radius. This temperature depends only on the cluster heating 
efficiency ($\eta$). Based on our  two dimensional simulations (Wunsch et al. 
2008) we calculate  the line profiles that result from several models 
and confirm our analytical predictions. From a comparison between the 
predicted and observed values of the half-width zero intensity of the two 
line components we conclude that the thermalization efficiency in SSC's 
above the threshold line must be lower than 20 \%.
\end{abstract}


\section{Introduction}

Young super star clusters (SSCs) with ages  less than 10$^7$ years, stellar 
masses $\sim 10^5$ to 10$^7$ M$_\odot$, and sizes which span only a few 
parsecs have been found  in a large variety of galaxies. Many of these  
have turned out to be strong line emitters in the infrared and radio regimes
(see, for example, Turner et al. 2000; Gilbert et al. 2000; Galliano \& 
Alloin 2008).
Some of them, as in the Antennae galaxy, Henize 2-10, NGC 5253 and IIZw 40 
(see Gilbert \& Graham 2007; Henry et al. 2007; Turner et al. 2003; Beck et 
al. 2002) are known to emit spatially extended Br$\gamma$ lines from dense 
HII regions. Most of these present supersonic line widths (FWHM $\sim$ 50 to 
200 km s$^{-1}$) that may exceed the escape velocity of the clusters. As 
reported by Beck (2008), in some cases the emission lines can be fitted by a 
single gaussian component, as it is the case of NGC 5253, while others (as in 
He 2 - 10) require also of a low intensity, although much broader component 
(FWHM $\sim$ 250 - 550 km s$^{-1}$), to fit the observed lines. A central issue 
pointed out by Turner et al. (2003) when dealing with the central supernebula in NGC 5253 which presents a mean radius 
smaller than 1 pc, is that  
 if the supersonic line width is caused by the expansion of the  nebula,
then its dynamical age is implausibly short. 
The supersonic infrared (Turner et al. 2003; Henry et al 2007) and radio 
recombination lines (Rodr\'\i guez-Rico et al. 2007) lack a full explanation.
The core high intensity, broad component has been interpreted 
as arising from the photoionized gas leftover from the process of star 
formation, while the low intensity and even broader component may be the 
signature of an outflow, a cluster wind, that begins to disrupt the leftover 
cloud and unveil the central SSC. These ideas are supported by the high 
extinction and the estimated high  density gas ($n_{HII} \sim 10^4 - 10^5$ 
cm$^{-3}$) in the observed HII regions, which may 
indicate that the star-forming episodes are young. Another possibility in the 
literature comes from an analogy to the broad lines presented by galactic 
compact HII regions. Henry et al (2007) and Beck (2008) have suggested that 
a plethora of massive stars blowing their winds may also contribute to the 
core line widths. Normal O stars however have winds with velocities 
$\sim$ 10$^3$ km s$^{-1}$, but  in their view, a dense environment may reduce the speed to the observed values, as it is the case in compact HII regions in our galaxy (De Pree et al. 2000).

Here we take a completely different approach based on the recent theoretical 
contributions (Silich et al. 2004, 2007; Tenorio-Tagle et al. 2005, 2007; 
W\"unsch et al. 2007, 2008) that have led us to realize the physical 
conditions that prevail within the volume occupied by the large collection of 
massive stars expected within young and compact SSCs. The model assumes, as 
in the original contribution of Chevalier \& Clegg (1985), that there is an 
even distribution of massive stars within the cluster volume, and considers 
their strong winds as well as their mass and energy released through 
supernova events. It considers also the direct interaction of all of these 
supersonic streams and thus the thermalization of the kinetic energy 
provided by massive stars. Our approach takes into consideration radiative 
cooling in the thermalized plasma and the thermalization efficiency $\eta$ 
which accounts for the immediate  loss of energy resultant from the proximity 
of the sources and the large metallicities of the reinserted matter. Thus 
$\eta$, the thermalization efficiency, measures the remaining energy available
to the matter reinserted into the cluster volume. We neglect the radiation
pressure as a possible wind driving mechanism because in a typical medium inside the young SSC's the dust
grains are destructed by sputtering on a very short time scale. Moreover, the effect of radiation is
small compared to the thermal pressure driving (see the Appendix).

Here we show that the detected broad nebular lines have indeed nothing to do 
with the expansion of the nebula. The supersonic broad emission lines  result 
only in clusters undergoing the  bimodal hydrodynamic solution (see
 Tenorio-Tagle et al. 2007),  those  found above the threshold 
line as defined in the  cluster mass or mechanical energy vs size plane by Silich et 
al. (2004).  The origin of the  most intense of the  broad lines is due to a plethora of  
repressurizing shocks (see, for instance, Zeldovich \& Raizer, 1965), 
induced within the  dense thermally unstable reinserted gas as this strives to 
maintain pressure balance with the much hotter gaseous counterpart. The less 
intense, although much broader gaussian component detected  only in some cases, is 
here shown to be  caused by the cluster wind,  as it becomes photoionized and less dense upon its own expansion.
Section 2 summarizes the hydrodynamics  of the matter reinserted within SSCs.
Section 3 provides a handle  on the line-widths expected for clusters 
undergoing  the bimodal solution and shows that these are a measure of the 
heating efficiency attained by the matter reinserted within SSCs. Section 3 
provides also some line profile examples calculated from our two dimensional hydrodynamic 
simulations (W\"unsch et al. 2008) and section 4 summarizes our conclusions.

\section{The state of the matter reinserted within SSCs}

Within young SSCs, the continuous mass and 
energy deposition, due to stellar winds and frequent supernova explosions, 
rapidly builds up extreme conditions. These favor either a strong stationary 
wind, the cluster wind, for clusters below the threshold line in the mass or 
mechanical luminosity vs size plane, or a bimodal flow for 
clusters above the threshold line (see Silich et al. 2004; Tenorio-Tagle et 
al. 2007). In either of these 
possibilities, the reinserted kinetic energy ($L_{SC}$) is immediately 
thermalized across the multiple shocks that result from the interaction of 
neighboring supersonic streams. This also leads to a high temperature 
($T_{SC} \sim 10^7$ K) plasma, and in the first possibility (below the
threshold line), to a high pressure environment, that powers its own 
expansion  into the low pressure surrounding medium. In this way a stationary 
wind ($\dot M_{SC}$ = $ 4 \pi R_{SC}^2 \rho_{SC} c_{SC}$; where 
$\dot M_{SC}$ is the mass deposition rate inside the star cluster volume and 
$\rho_{SC}$ and $c_{SC}$ are 
the density and sound speed at the cluster surface $R_{SC}$) is established.  
The wind is to become strongly radiative, departing from the adiabatic 
solution of Chevalier \& Clegg (1985), for clusters near  the threshold line 
(see Silich et al. 2004). 
In the bimodal cases (ie. above the threshold line),
radiative cooling depletes rapidly the energy 
gained through thermalization, particularly within parcels of gas in the 
densest central regions of the cluster. As shown by Tenorio-Tagle et al. (2007)
and by W\"unsch et al. (2007, 2008), radiative cooling  forces the stagnation 
radius $R_{st}$ (the radius where the velocity of the flow equals 0 km 
s$^{-1}$) to move out of the cluster center, while driving  the matter
reinserted within this inner volume to frequently become thermally unstable. 
At the same time, the matter reinserted between $R_{st}$ and the cluster 
surface, although also exposed to strong radiative cooling, manages to 
compose a stationary wind.
  
Thus in both cases, below and above the threshold line, the cluster stars end 
up immersed in a pervasive hot medium. Below the threshold line this streams 
out as a  hot ($T$ $\sim$ 10$^7$ K) cluster wind as it is inserted.  Such a wind will only cool down 
at large distances (tens of pc) from the cluster surface (see Silich et al. 2004) and thus it would not lead to a broad line component associated to the star cluster. Note that the observed spectra are obtained by integrating over less than $1.5^{\prime\prime}$ along the slit, centered on the emission peaks, and thus the observed broad lines are indeed associated to the SSCs. In the bimodal cases however, pockets
of gas within the volume enclosed by the stagnation radius, are to become 
frequently thermally unstable. Radiative cooling brings rapidly their 
temperature down to 10$^4$ K or even to lower values, if an ionizing photon 
flux is not sufficient to fully ionize the resultant condensations
(say, for coeval clusters with an age $t  > 10^7$ yr). As shown by 
W\"unsch et al. (2008), in all bimodal cases, most of the dense condensations 
generated within the stagnation volume are unable to leave the 
cluster and thus inevitably accumulate within the cluster volume.  
Eventually such a process results into further generations of stellar 
formation (Tenorio-Tagle et al. 2005; Palou\v{s} et al. 2008). 

In the bimodal cases, the sudden loss of temperature and thus pressure, 
immediately causes the appearance of strong repressurizing shocks 
(Zeldovich \& Raizer, 1965; Shapiro \& Kang, 1987; Vietri \& Pesce, 1995;
Tenorio-Tagle, 1996). 
These emanate from the pervasive hot, high pressure gas and move with large 
speeds into the thermally unstable parcels of gas, causing their rapid 
condensation through a strong reduction of their volume and  the corresponding
density enhancement. Note that the density (or metallicity) of the unstable 
gas needs to be only slightly larger than the density (or metallicity) of the 
stable counterpart for the instability to occur and thus, one can show that 
the asymptotic speed of the repressurizing shocks under the adiabatic
($V_{RA}$) and isothermal ($V_{RT}$) approximations is only a 
function of the temperature ($T$) of the hot gas (Vietri \& Pesce, 1995): 
\begin{eqnarray}
      \label{eq1a}
      & & \hspace{-1.0cm}
V_{RA} = \alpha_A \left(\frac{\gamma + 1}{2} \frac{k T}{\mu m_p}\right)^{1/2} 
      \\[0.2cm]
      \label{eq1b}
      & & \hspace{-1.0cm}
V_{RT} = \alpha_T \left(\frac{k T}{\mu m_p}\right)^{1/2} , 
\end{eqnarray}
where $k$, $\mu$ and $m_p$ are the Boltzmann constant, the mean mass per 
particle in the ambient hot plasma and the proton mass, respectively. The 
parameters $\alpha_A$ and $\alpha_T$ are of the order of unity and their 
values depend on detailed physics of the highly supersonic repressurizing 
shocks including the geometry of the dense low pressure regions. They must
be determined via numerical simulations.
Thus, for the bimodal cases, the radiative wind theory leads to a plethora 
of cooling condensations, appearing in a recurrent frequent manner within 
the stagnation
volume. These are squeezed by multiple repressurizing shocks, driven 
into all sides of the thermally unstable parcels of gas by the higher ambient
pressure. At the same time the outer cluster regions (between $R_{st}$ and 
$R_{SC}$) compose a stationary wind, also exposed to strong radiative cooling 
and thus to be photoionized as it streams away from the cluster, when its  
temperature drops below $10^5$~K (Silich et al. 2004, 2007).       
However, in many cases the wind with the appropriate  temperature may present a  low density and thus lead to a very weak component, compared to the line produced by the dense gas behind the repressurizing shocks.

\section{Properties of the thermalized plasma}

The density and temperature distribution of the thermalized reinserted 
matter is such that it causes a slight outward pressure gradient and this 
leads to a smooth acceleration from a subsonic expansion, within the cluster 
volume, to a high velocity, supersonic flow outside of the cluster 
(Chevalier \& Clegg, 1985; Canto´ et al. 2000; Silich et al. 2004). This 
requires that the sonic point (the point where the sound speed is equal to the 
local expansion velocity) is located at the star cluster surface. One 
can always fulfil this condition by iterating over the temperature $T_{st}$ 
at the stagnation radius $R_{st}$. In turn, $T_{st}$ is coupled to the density 
$n_{st}$ and pressure $P_{st}$ at the stagnation point, and thus completely 
defines the thermodynamics of the re-inserted plasma there 
(see Silich et al. 2004; W\"unsch et al. 2007): 
\begin{equation}\label{eq3}
n_{st} =  q_m^{1/2} \left[ \frac{V_{A\infty}/2 - c^2_{st} /(\gamma - 1)}
{\Lambda (T_{st} , Z)}\right]^{1/2},
\end{equation}
\begin{equation}\label{eq4}
P_{st} = kq_m^{1/2} T_{st}
\left[\frac{V_{A\infty} /2 - c^2_{st}/(\gamma - 1)}{\Lambda(T_{st}, 
Z)}\right]^{1/2}
\end{equation}
where $c_{st}$ is the sound speed at $R_{st}$, $q_m$ is the mass deposition 
rate per unit volume, $V_{A\infty} = (2\eta L_{SC} /\dot{M}_{SC} )^{1/2}$ is the 
adiabatic wind terminal speed, and $\Lambda (Tst , Z)$ is the cooling 
function at the stagnation point. In this parameter space the value of 
$T_{st}$ depends only on one parameter, $\eta$, which in the semi-analytic 
models accounts for all uncertainties dealing with the thermalization of 
the kinetic energy provided by massive stars inside the star cluster volume
(see, for instance, W\"unsch et al. 2007; Silich et al. 2007). 
It defines the fraction of the star cluster mechanical luminosity that 
remains as  thermal energy of the re-inserted matter after strong radiative 
cooling removes during thermalization a fraction of the deposited energy. 
This implies that the amount of thermal energy deposited into the star 
cluster volume per 
unit time ($\eta L_{SC}$) is smaller than the total star cluster mechanical 
luminosity, $L_{SC}$, with  $\eta < 1$. The main impact of $\eta$ 
is to lower the location of the threshold line in the $L_{SC}$ vs $R_{SC}$
plane. 

The radiative solution found by Silich et al. (2004), Tenorio-Tagle et al. 
(2007), W\"unsch et al. (2007, 2008) defines the 
location of the  threshold line $L_{crit}(R_{SC})$, in the $L_{SC}$ vs $R_{SC}$
plane, which separates clusters (below the threshold line) for which
$R_{st}$ is at the cluster center, from those (above the threshold line)
with $R_{st} > 0$~pc (Tenorio-Tagle et al. 2007; W\"unsch et al. 2008). For 
clusters whose mechanical luminosity 
is equal or larger than the critical value, the temperature at the stagnation 
radius is defined by the condition that $P_{st}$ reaches its maximum possible 
value ($\dif{P_{st}}/\dif{T_{st}} = 0$) and this is  independent  of the star 
cluster mass and radius (Tenorio-Tagle et al. 2007).
\begin{figure}[htbp]
\plotone{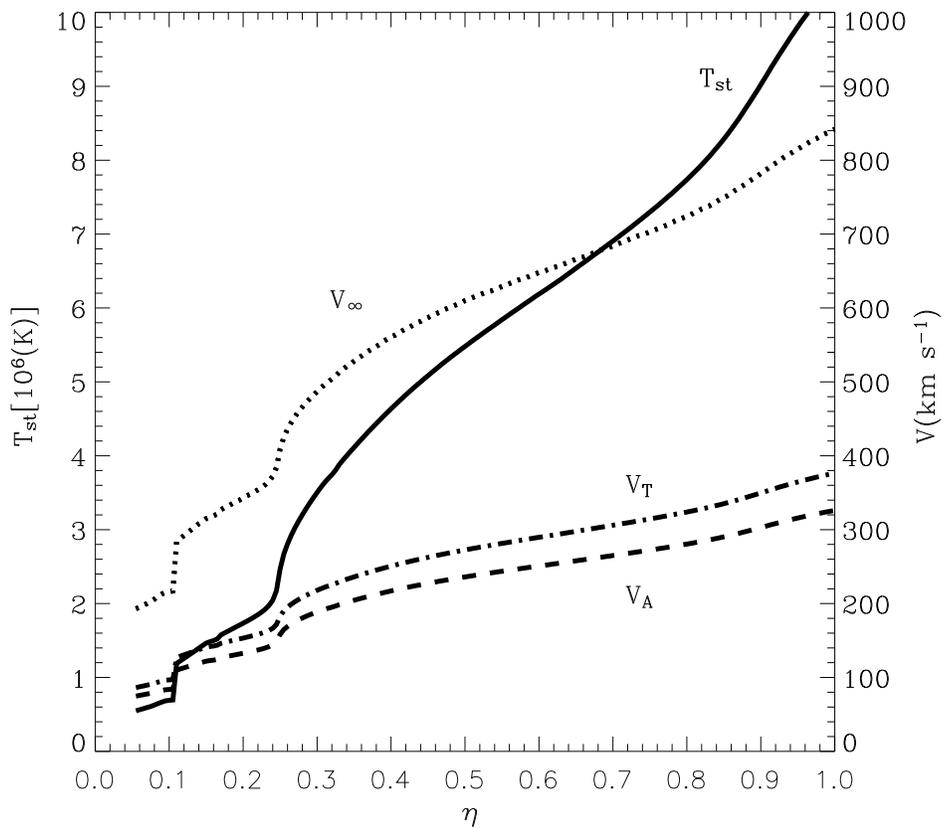}
\caption{The  velocity of the gas overtaken by the repressurizing  shock as 
a function of heating efficiency. The solid line marks the temperature at the 
stagnation radius ($T_{st}$), the dashed lines indicate the velocity of the 
gas overtaken by the repressurizing shocks under the adiabatic (\ref{eq1a}) 
and isothermal (\ref{eq1b}) approximation and the dotted line the terminal 
speed of the cluster wind ($V_\infty$), all of them plotted as a function of 
the heating efficiency, $\eta$.}
\label{fig1}
\end{figure}

For massive and compact star clusters, which are located above the 
threshold line in the $L_{SC}$ vs $R_{SC}$ plane, one can present
the condition $\dif{P_{st}}/\dif{T_{st}} = 0$ in the form of a nonlinear 
algebraic equation: 
\begin{equation}
      \label{eq5}
\left(\frac{\eta V^2_{A\infty}}{2} - \frac{c^2_{st}}{\gamma - 1}\right)
\left(1 - \frac{T_{st}}{2 \Lambda} \der{\Lambda}{T_{st}}\right) -
\frac{1}{2} \frac{c^2_{st}}{\gamma - 1} = 0 .
\end{equation}
One can then use this equation in order to calculate the value of $T_{st}$ 
for different values of the heating efficiency $\eta$.
Note that the temperature of the thermalized plasma remains almost constant 
inside the stagnation radius ($T(r < R_{st}) \approx T_{st}$), and thus 
$\eta$ is here used to define through equations (\ref{eq1a}) and (\ref{eq1b})
the expected asymptotic velocity of the repressurizing shocks and the 
terminal wind velocity (Silich et al. 2007):
\begin{equation}
      \label{eq6}
V_{\infty} = \left(\frac{2}{\gamma-1}\right)^{1/2} c_{st} ,
\end{equation}
Figure 1 displays the values of $T_{st}$, $V_{\infty}$ and the 
velocity of the gas overtaken by the repressurizing shocks 
($V_{A} = 0.75 V_{RA}$ in the adiabatic and $V_{T} = V_{RT}$ in the 
isothermal case) calculated for different values of the heating efficiency by 
means of equations (\ref{eq1a}), (\ref{eq1b}) and (\ref{eq6}) under the
assumption that $\alpha_A = \alpha_T = 1$.

\subsection{The broadening of the nebular  lines}

In the context of the matter cooling within the superstar cluster stagnation
radius, the repressurizing  shocks have been calculated first by means of 1D
numerical hydrodynamics in Tenorio-Tagle et al. (2007) and in two dimensional
calculations by  W\"unsch et al. (2008). We use the later to calculate
recombination line profiles and present results for three models with the 
following
parameters: Model 5 ($R_\mathrm{SC} = 10$~pc, $L_\mathrm{SC}/L_\mathrm{crit} =
20$ and $\eta = 1$), Model 6 ($R_\mathrm{SC} = 10$~pc,
$L_\mathrm{SC}/L_\mathrm{crit} = 200$ and $\eta = 1$) and Model 8
($R_\mathrm{SC} = 3$~pc, $L_\mathrm{SC}/L_\mathrm{crit} = 2.5$ and $\eta =
0.3$); see W\"unsch et al. (2008) for further details.
\begin{figure}[htbp]
\plottwo{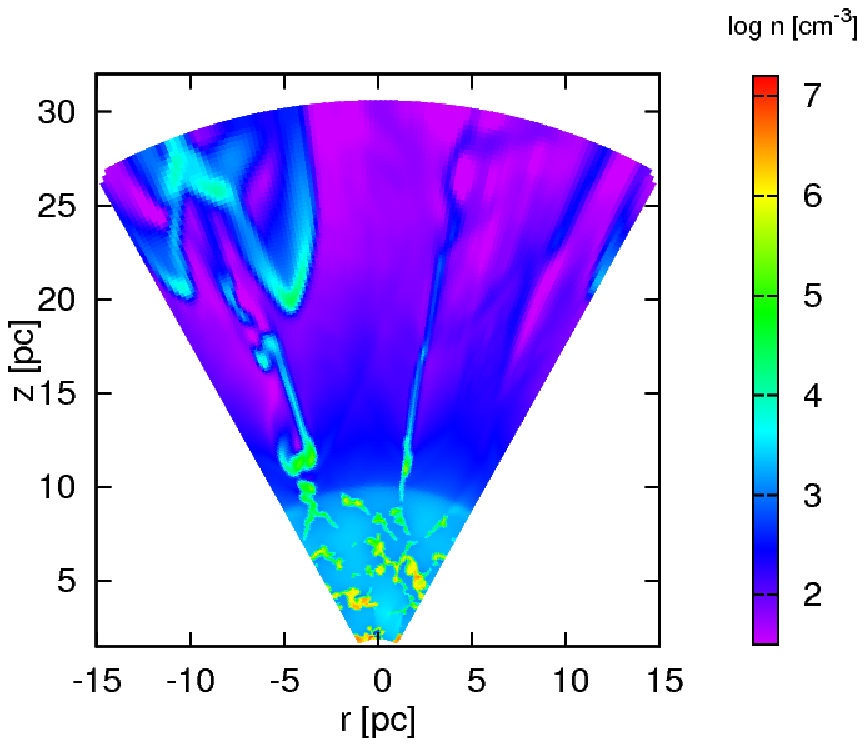}{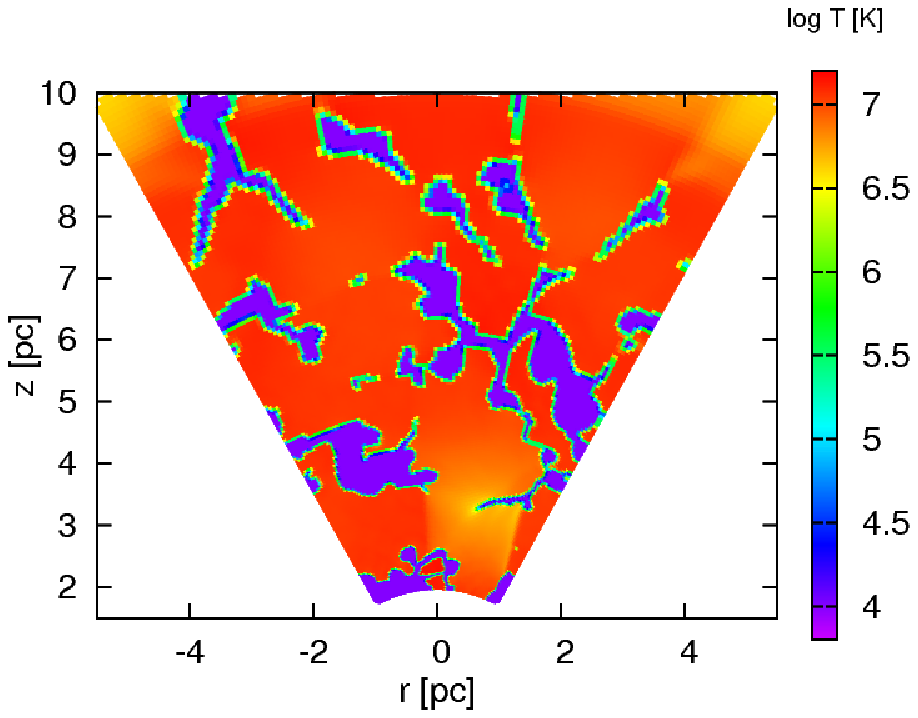}
\caption{Model 5. Distribution of particle density (left) across the whole
computational domain and temperature (right) inside the cluster at time
$0.345$~Myr from the beginning of the simulation.}
\label{m5_CD}
\end{figure}

The simulations were carried out on a spherical ($r$,$\theta$) grid consisting
of $N_i\times N_j$ elements with a typical extent in the $\theta$-direction
$\pi/3$ (see Figure~\ref{m5_CD} which shows the logarithm of the particle 
density and
temperature in one of the simulations). To obtain the line profile, we extend
the computational domain into a three dimensional sphere by copying and 
rotating
it twice in the $\theta$-direction and $N_k-1$ times in the $\phi$-direction. The
resulting sphere consists of $N_i\times 3N_j\times N_k$ elements where indices
$i, j$ and $k$ identify the position of element in the $r, \theta$ and $\phi$
directions, respectively. Since the $\phi$-component of velocity does not exist
in 2D simulations, we assume that the two axial directions are statistically
equivalent and rotate the $\theta$-component of the velocity by a random angle
in the $\phi$-direction by setting $v_{\phi,ijk} = v_{\theta,ijk}^\mathrm{2D} \sin\alpha$
and $v_{\theta,ijk} = v_{\theta,ijk}^\mathrm{2D} \cos\alpha$ where $\alpha$ is a random
number between $0$ and $2\pi$.

Only elements with $5 > \log T_{ijk}/[K] > 4.3$ and elements with $\log
T_{ijk}/[K] \le 4.3$ and $n_{ijk} < n_\mathrm{lim}$ are taken into 
consideration
for the computation of the line profiles. The high temperature limit, 
$T_\mathrm{max} = 10^5$~K, ensures that only recombining gas contributes to 
the line. The low temperature limit, $T_\mathrm{min} = 10^{4.3}$~K, is used 
to exclude the gas in dense warm condensations whose interiors would cool 
even further and collapse into 
stars. However in the simulations we artificially maintain the clump 
temperature at $10^4$~K. This procedure was introduced because the code does 
not include transport of ionizing radiation which would allow to determine 
properly the temperature across the dense condensations. We do not apply the 
low temperature limit
for densities below $n_\mathrm{lim}$ (different for simulations with different
cluster parameters), because we would exclude the rarefied wind which cools 
down as it flows out of the cluster. We set $n_\mathrm{lim} =
300$, $300$ and $10$~cm$^{-3}$ for Model 5, 6 and 8, respectively.
Figure 3 (left panels) shows individual grid 
cells of the 2D computational domain as points in $\log n$ vs $\log T$ plane 
and the color denotes the magnitude of the velocity in the cell. Regions 
contributing to the line calculation are enclosed by the dashed lines. 
The figures do show the broad range of physical conditions within the matter associated to SSCs.
This goes from the densest and coolest gas to the hottest recently re-inserted matter, some of which cools rapidly  as it streams as a wind. 

For each element, we determine the line-of-sight velocity
$v_{\mathrm{los},ijk}$ and the luminosity
\begin{equation}
\mathrm{L}_{ijk} \sim \rho_{ijk}^2 V_{ijk} ,
\end{equation}
where $\rho_{ijk}$ and $V_{ijk}$ is the mass density and volume of the element,
respectively. We divide the range of all possible line-of-sight velocities into
bins of sizes $dv_\mathrm{los} = 10$~km\,s$^{-1}$ and add together contributions
of all elements with $v_{\mathrm{los},ijk}$ in a given bin. Finally, we convolve
the calculated line profiles with a Maxwell velocity distribution function for
$c_s = 10$~km\,s$^{-1}$ and normalize the result to the total line luminosity.

Line profiles computed from cells with $r < R_\mathrm{SC}$ (red), 
$r > R_\mathrm{SC}$ (green) and their sum (blue) are shown in the
right panels of Figure 3. 
The half width zero intensity (HWZI), which corresponds
to the maximum velocity of the gas, of the narrow component, coming from the
cluster interior, is in a good agreement with velocities predicted by Equations
(\ref{eq1a}) and (\ref{eq1b}) for a given $\eta$. Similarly, the HWZI of the 
broad component, coming from outside of the cluster, is in a good agreement 
with the velocity predicted for the terminal velocity of the cluster wind
(equation \ref{eq6}).

Finally, we have checked how the line profiles depend on values of
$T_\mathrm{min}$ and $n_\mathrm{lim}$. It turns out that a ratio of intensities
of the two line components is really sensitive to them. It is because by
decreasing $T_\mathrm{min}$ and increasing $n_\mathrm{lim}$, more and more dense gas with
preferentially smaller velocities is included into the line computation.
However, the HWZI of the two components remains always the same.
Therefore, we conclude that 2D simulations lead to supersonic emission lines with
two (narrow and broad) components  with  widths in agreement with our analytical
results for repressurizing shocks and the cluster wind, respectively. 

We thus associate the supersonic line profile produced by the multiple
repressurizing shocks that evolve within the stagnation volume to the high
intensity central component detected in massive and compact SSCs. On the other
hand, the cluster wind, promotes the appearance  of a broader low intensity
component. As shown in Figure 1, both speeds, that of the repressurizing shocks
and that of the cluster wind, attain a unique value for every $\eta$, and these
should correspond to the largest speeds (HWZI) that one could infer from the
gaussian components used to fit the observed lines.  
This is the case for example in He 2-10. Henry et al. (2007) display the two Br$\gamma$ gaussian components 
for several sources in their Figure 5. If one measures the HWZI of both of these components, both of them independently lead, through Figure 1, to the same value of $\eta$. We note than in all of these cases 
the resultant value of $\eta$ is below 20 percent. 
To fit the two gaussians is sometimes a complicated issue as in Figure 6 of Henry et al. However, if one measures the HWZI of the observed line and assumes that this largest detected speed is due to the wind emanating from the cluster, then Figure 1 will indicate what the value of $\eta$ is. Our findings thus provide a unique tool to measure directly the heating efficiency 
of SSCs, otherwise, a very elusive quantity.

\section {Discussion}

The supersonic broad nebular lines detected within young, massive and compact 
SSCs have been interpreted as arising from the photoionized gas left over 
from the process of star formation. Despite the fact that such a claim 
lacks a thorough proof, one should notice in this respect that in the 
bimodal cases the rapid return of stellar matter into the cluster volume 
is to lead, in just a few $10^6$ years, to the large densities ($n_{HII} 
\sim 10^4 - 10^5$ cm$^{-3}$) detected in these objects.
  
We have shown here that for clusters undergoing a bimodal hydrodynamic 
solution, those located in the $L_{SC}$ vs $R_{SC}$ plane above the threshold 
line, the radiative wind model predicts, despite the large energy input rate,  
the recurrent appearance of high density condensations continuously 
accumulating within the stagnation volume. The condensations result from 
parcels of gas that become thermally unstable and thus, having lost their 
large temperature ($\sim T_{st}$) are then compressed by a plethora of 
repressurizing shocks while being subjected to become photoionized 
by the cluster radiation field, conforming an HII region right within the 
cluster volume. Thus, the bimodal model predicts a hot thermalized plasma 
that co-exists with the warm photoionized gas and the resultant cold dense 
condensations.  

Furthermore, the model  predicts two sources of broadening for the nebular 
lines: the repressurizing shocks and the cluster wind. We associate the 
denser and slower of these, resultant from the plethora of repressurizing 
shocks, to the supersonic central component detected in most objects, while 
the much broader and less intense lines, should emanate from the radiative 
cluster wind.  There is an excellent agreement between the half width zero 
intensity of the broad lines that result from our 2D calculations and the 
analytic formulae (equations \ref{eq1b} and \ref{eq6}) derived for the 
isothermal velocity of the repressurizing shocks and the wind terminal speed
both here shown to depend only on $\eta$.
Thus in our model the line broadening has nothing to do with 
the expansion of the nebula or with the gravity of the system, it is largely 
due to internal motions repeatedly appearing as the reinserted matter becomes 
thermally unstable and to the cluster wind photoionized near the cluster 
surface. Here we have also shown that the broadening of the nebular
lines provides an accurate method to infer the heating efficiency within SSCs.
The FWHM values of 50 - 200 km s$^{-1}$ for the narrow and 250 - 550 km 
s$^{-1}$ for the broad line profile components used to fit the line emission 
from  clusters reported as presenting supersonic line broadening (see 
Section 1) imply low values of the heating efficiency, $\eta$, smaller that 
20 percent. Another independent determination of $\eta$ (Silich et al. 2007;
2009) predicts lower but similar values (smaller than 0.1) for M82-A1 
and several other clusters in the central zone of M82.

The values $T_{min}$ and $n_{lim}$ of the lower bound in the temperature
interval and the upper bound in density considered in the line profile 
construction are  linked to the thickness
of the layer on  the surface of warm dense clumps from where the lines
emerge. To justify the choice of an specific value, we would need to
perform the radiative transfer calculations inside of warm clumps and
subsequent processes leading to star formation. This will be  important
for the discussions of the exact line profile  shapes. However, here our 
aim, our interest, is the terminal velocity of the two line components,
which, as shown, depend only on the parameter $\eta$.
We expect the repressurizing shocks to lead to the most intense line.
This is because of the large densities involved. The line produced by the 
wind is always due to  be weaker although broader. And thus if a source 
presents only  one component, this has to  emanate from the interior of the 
cluster,  from  the matter behind the repressurizing shocks.

There is an excellent agreement between the half width zero intensity
of the broad lines that result from our 2D calculations and the analytic 
formulae (equations \ref{eq1b} and \ref{eq6}) derived for the isothermal 
velocity of the repressurizing shocks and the wind terminal speed
both here shown to depend only on $\eta$.

Note that for clusters undergoing the bimodal solution, within their 
volume defined by 
the stagnation radius the hydrodynamic evolution becomes rapidly  strongly 
coupled to the UV radiation field generated by the massive stars in the 
cluster. These through photoionization, keep the temperature of the thermally 
unstable gas at $T \sim 10^4$ K. This is particularly true for very young 
clusters, before the supernova era starts 
($t_{SN} \sim 3$~Myr). However, older clusters, with a reduced ionizing 
photon flux, soon  become unable to photoionize all the gas that has become 
thermally unstable. In this case, the thermally unstable gas would continue 
to cool further, while being compress into correspondingly smaller volumes.  
As a result, the increasingly higher densities would be able to trap the 
ionization front around the outer skins of the condensations while remaining 
neutral and at low temperatures ($\sim $ 10 K) within their cores. 
In this way, if a parcel of gas with an original temperature $\sim 10^7$ K, 
cools down to 10 K, then its volume, in order to preserve pressure 
equilibrium, would have been reduced six orders of magnitude while its 
density would become six orders of magnitude larger. 
Such clusters thus present matter with completely different physical 
conditions and thus should be easily observed in X-rays, Infrared and radio 
continuum, while presenting a large extinction at optical wavelengths.

\acknowledgments 
We thank our anonymous referee for valuable comments and suggestions.
This study has been supported by CONACYT - M\'exico, research grants 
60333 and 82912  and the Spanish Ministry of Science and Innovation under 
the collaboration ESTALLIDOS (grant AYA2007-67965-C03-01) and 
Consolider-Ingenio 2010 Program grant CSD2006-00070: First Science with the 
GTC. RW acknowledges support by the Human Resources and Mobility Programme 
of the European Community under contract MEIF-CT-2006-039802. RW and JP 
acknowledge support from the Institutional Research Plan AV0Z10030501 of the 
Academy of Sciences of the Czech Republic and project LC06014 Centre for 
Theoretical Astrophysics of the Ministry of Education, Youth and Sports of 
the Czech Republic. We also acknowledge CONACYT-Mexico \& the Academy 
of Sciences of the Czech Republic Grant 2009-2010.

\appendix

\section{On the effects of radiation pressure}

The effects of radiation pressure are not included since the dust evaporates 
rather fast due to sputtering in a medium with temperature $\sim 10^7$~K and 
density $\sim 10^3$~cm$^{-3}$. This may be estimated using a formula given by 
Silk \& Burke (1974) who gave the time-scale for destruction of dust grains 
by sputtering:
\begin{equation}
\tau_{sp} = 9.5\times 10^{6} \left(\frac{2.1\times 10^{-3} 
 \mathrm{cm}^{-3}}{n}\right)\left(\frac{1.4\times 
10^{8}~\mathrm{K}}{T}\right)^{1/2} \left(\frac{0.01}{Y}\right) \mathrm{yr}\ .
\end{equation}
Here $n$ is the number density, $T$ is temperature and $Y$ is sputtering
yield per collision, which is $\sim 0.01$ (see Figure 2, Burke \& Silk, 1974). 
This gives for our typical values about $30$~yr as the
grain destruction time-scale. Thus we conclude that there is little dust in our
wind, which brings the momentum delivery by radiation close to zero.
Another argument for not including dust is that we model a young 
cluster, where AGB stars, a potential major source of dust, are absent.

However, if the dust would be able to survive in the hot environment, 
the wind velocity may be estimated from the pressure on dust grains 
considering its opacity in the following way.
The equation of motion for a wind with optical depth $\tau$ inside  the 
star cluster volume is (see, for example, equation 24 in Murray et al. 2005):
\begin{equation}
       \label{eqAa}
u\der{u}{r} = - \frac{G M(r)}{r^2} + \frac{\kappa L_r(r) \exp(-\tau)}
                {4 \pi r^2 c} ,
\end{equation}
Outside  the star cluster volume it is:
\begin{equation}
       \label{eqAb}
u\der{u}{r} = - \frac{G M_{SC}}{r^2} + \frac{\kappa L_{bol} \exp(-\tau)}
                {4 \pi r^2 c} ,
\end{equation}
where $u$ is the flow velocity, $c$ is the speed of light, G  the  
gravitational constant, $M(r)$ is the mass within radius $r$, $L_{bol}$ is 
the star cluster bolometric luminosity, $L_r(r) = L_{bol} (r/R_{SC})^3$ and
$\kappa$ is the opacity per unit mass of gas. The optical depth is:
\begin{equation}
       \label{eq2}
\tau = \int_0^r \kappa \rho {\rm d} r .
\end{equation}
  
Let us neglect  the gravitational pull from the cluster. Then one can
integrate equations (\ref{eqAa}) and (\ref{eqAb}) taking
into consideration that ${\rm d}\tau = \kappa \rho {\rm d}r$, 
$L_{bol} = \epsilon L_{SC}$, where $L_{SC}$
is the star cluster mechanical luminosity and the maximum value of 
$\epsilon \approx 230$ according to the Starburst99 synthesis model 
(Leitherer et al. 1999), and that inside the cluster
\begin{equation}
       \label{eq3}
4 \pi r^2 \rho u = {\dot M}(r) = \frac{4 \pi}{3} q_m r^3 ,
\end{equation}
where  $q_m$ is the mass deposition rates per unit volume. The integration
yields:
\begin{equation}
       \label{eqA4}
u(\tau) = \frac{\epsilon V^2_{A\infty}}{2 c} (1 - \exp(-\tau)) .
\end{equation}

In our calculations it was adopted that $V_{A\infty} = (2 L_{SC} 
/ {\dot M}_{SC})^{1/2} = 1000$~km/s.
Equation (\ref{eqA4}) thus implies
that even if dust grains can survive inside the hot thermalized plasma, 
and if its optical depth is large ($\tau \rightarrow\infty$), the 
maximum flow velocity would be $\sim 380$~km\,s$^{-1}$ which is well 
below $V_{A\infty}$.

\newpage

\begin{figure}[htbp]
\plottwo{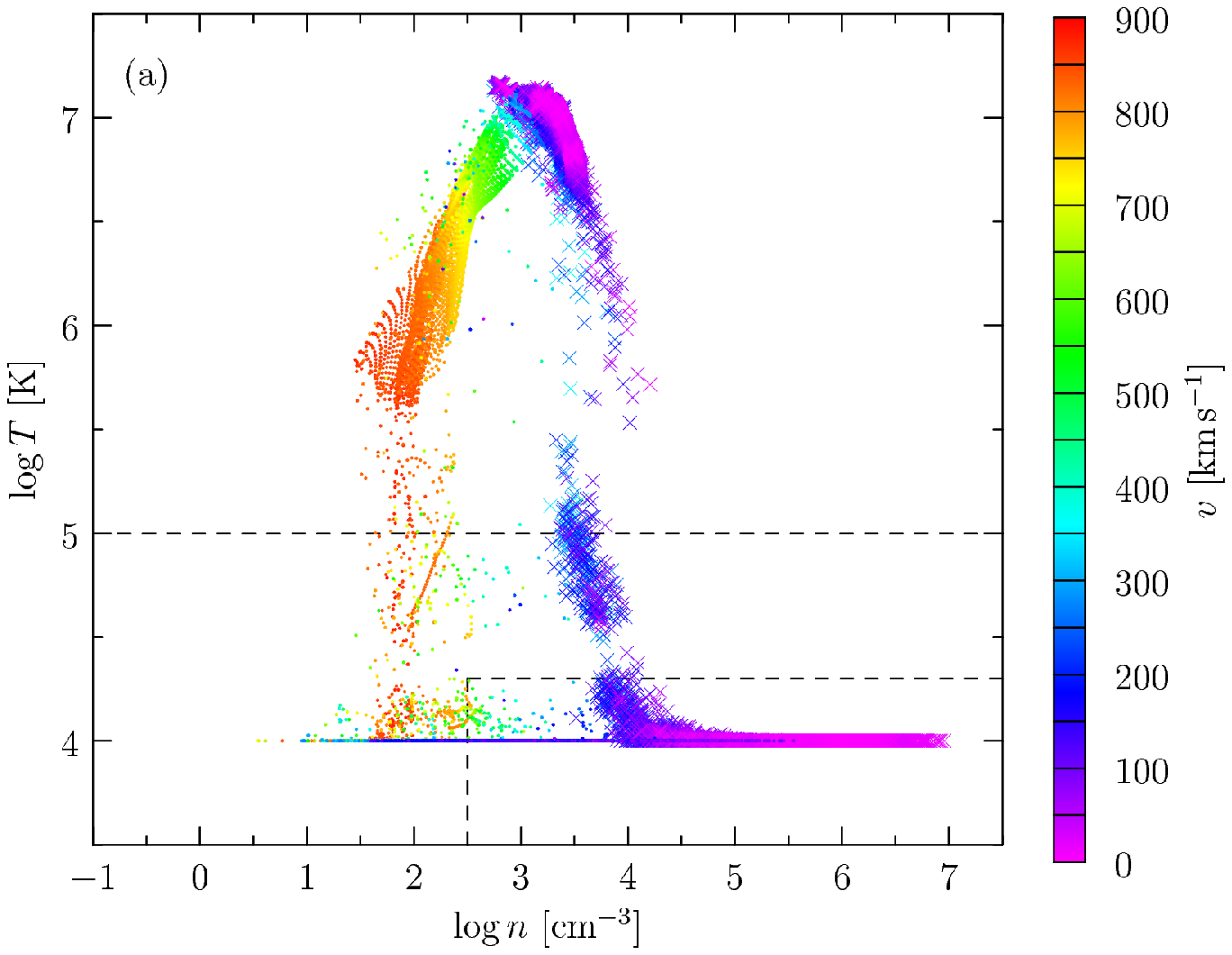}{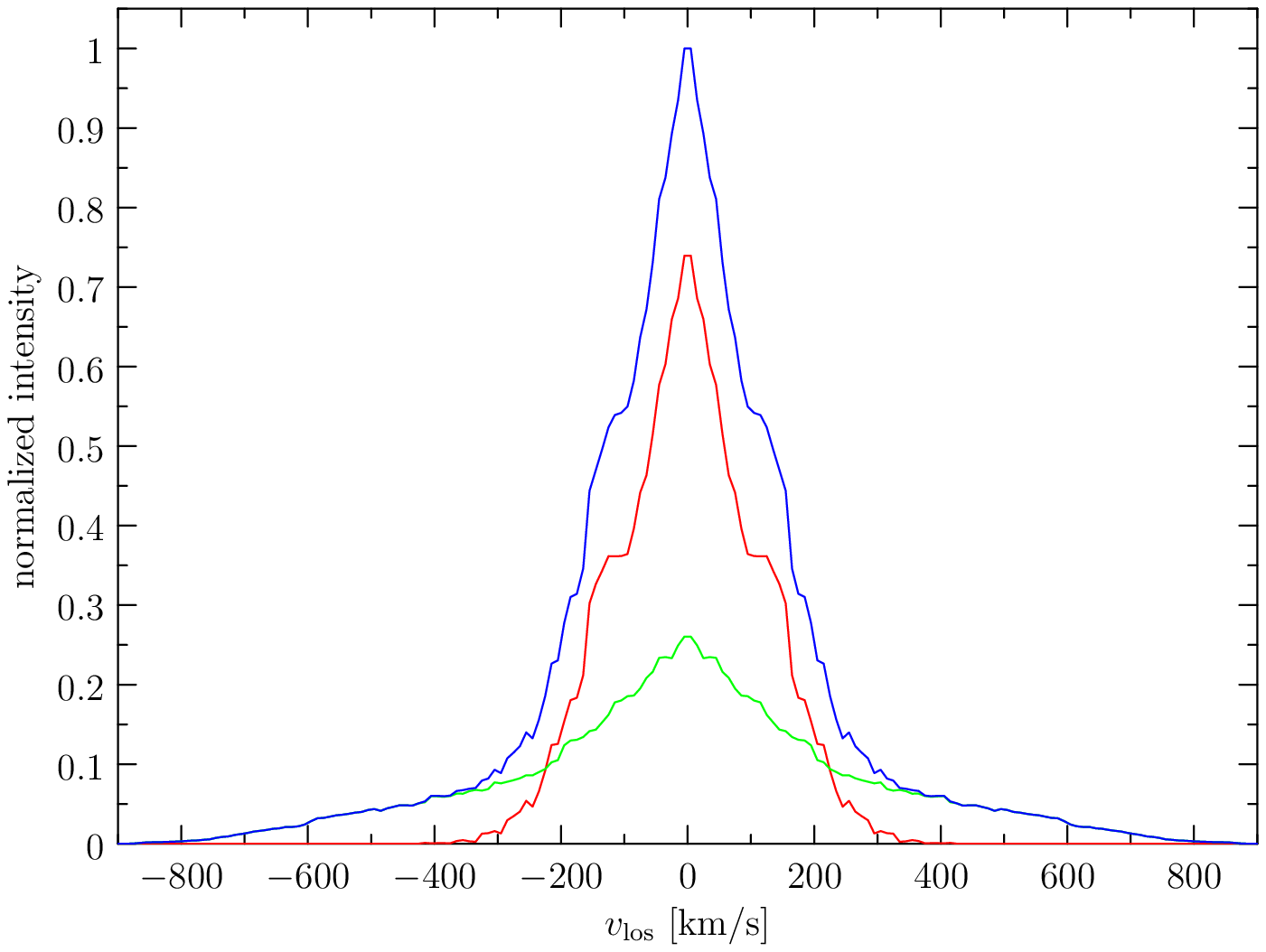}
\end{figure}
\begin{figure}[htbp]
\plottwo{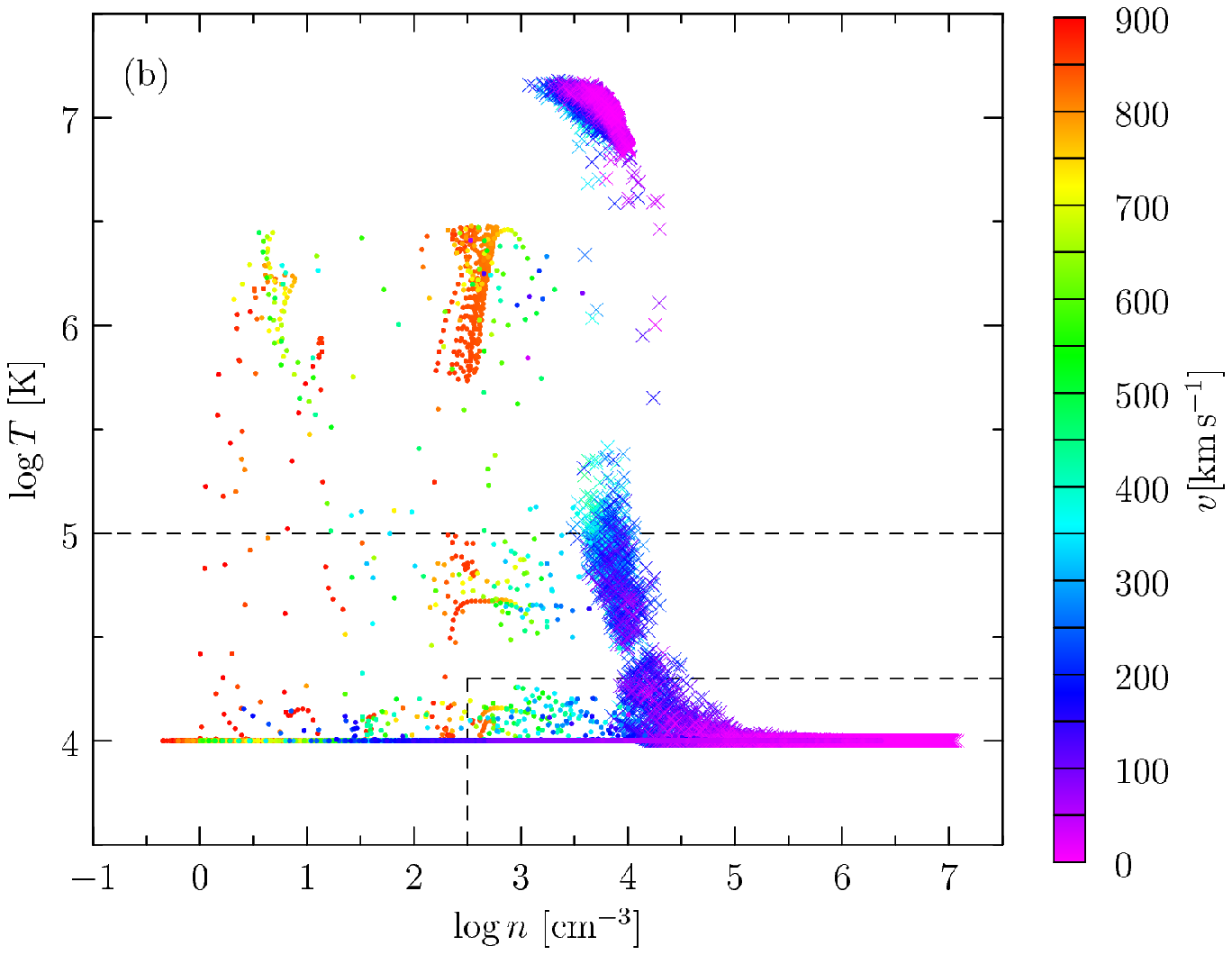}{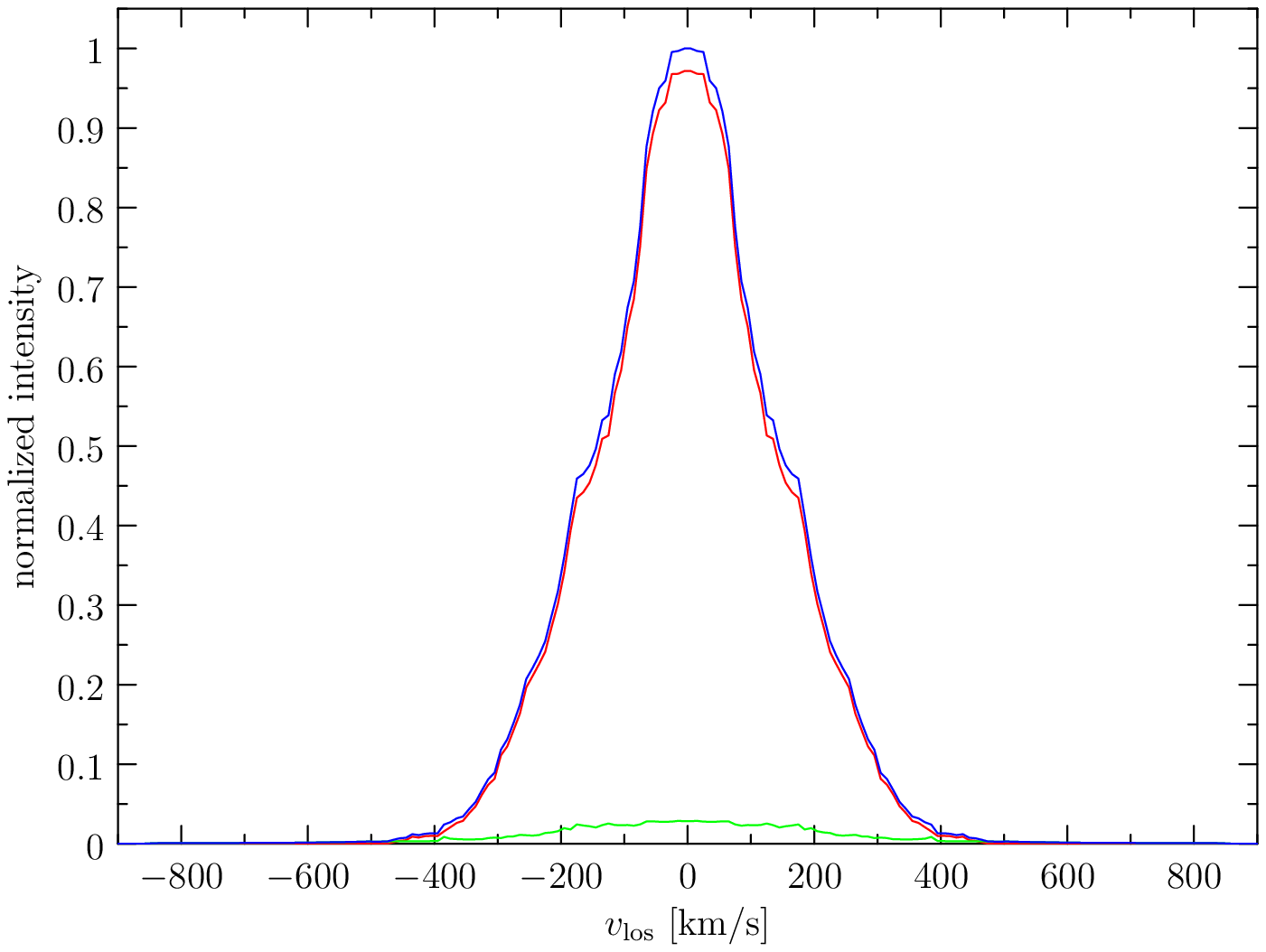}
\end{figure}
\begin{figure}[b]
\plottwo{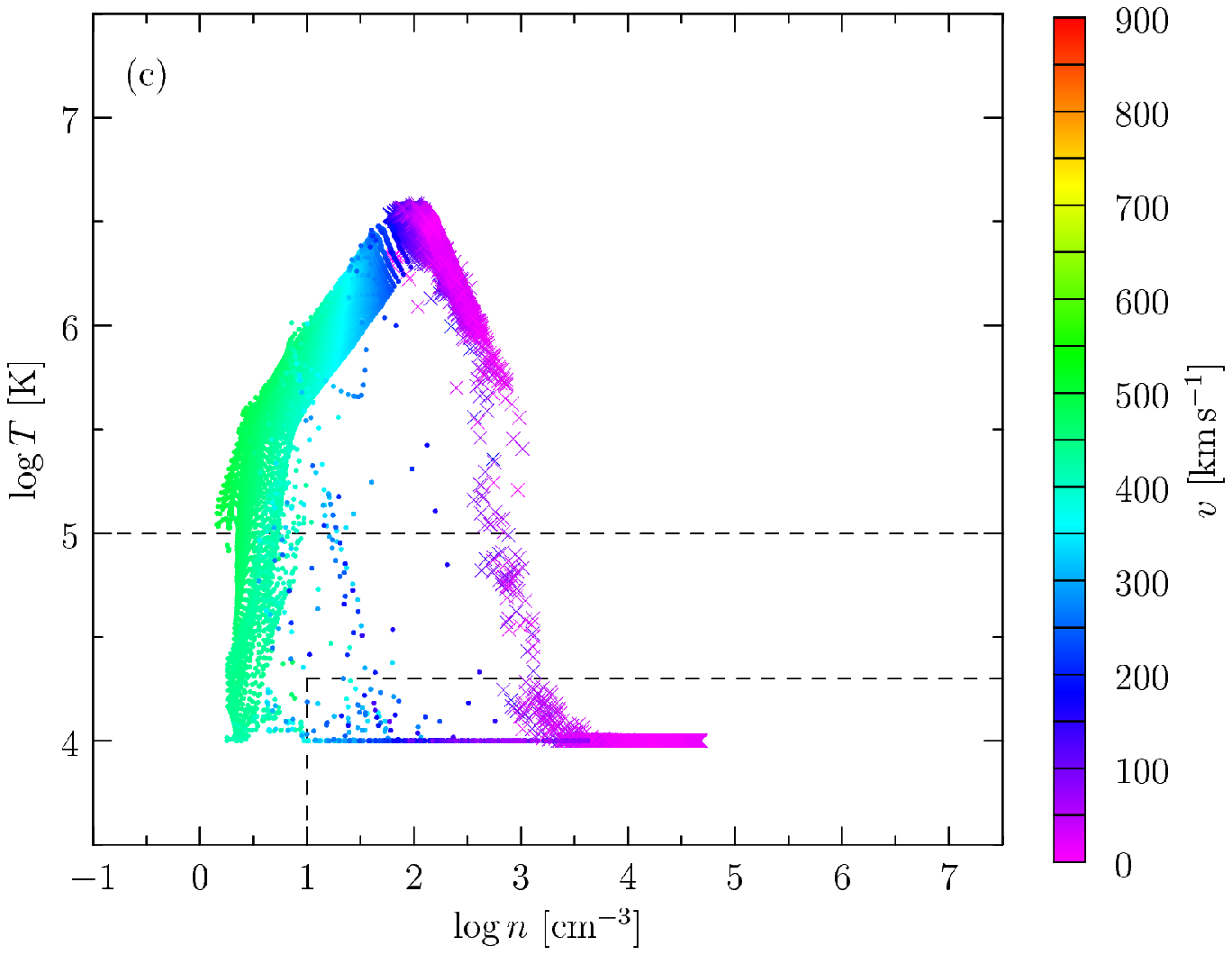}{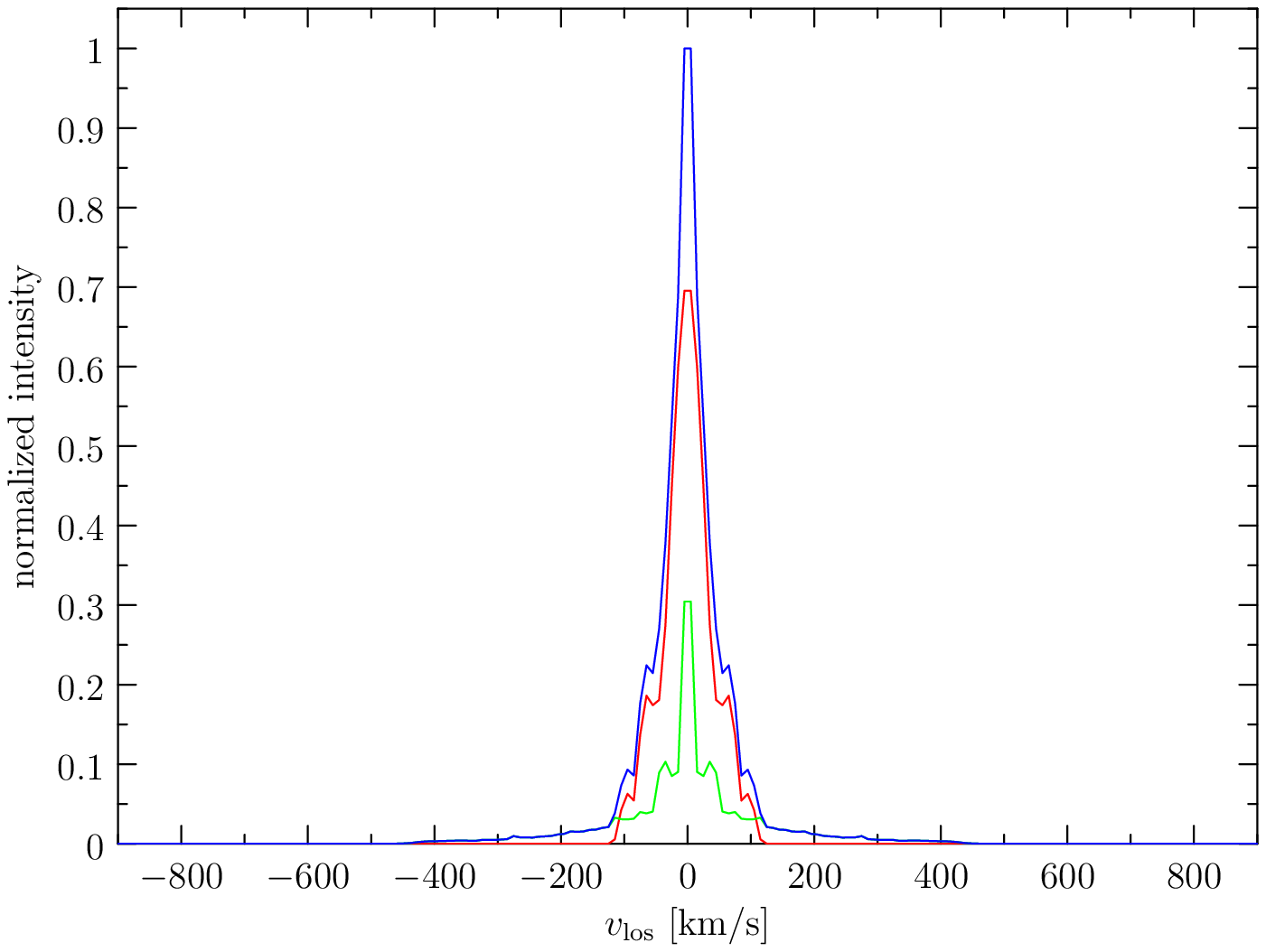}
\caption{Positions of simulation grid cells in the $\log n$ vs $\log T$ 
plane. 
Panels a, b and c present the results of the calculations for models 5, 6 and
8 at times $0.345$~Myr, $t = 0.355$~Myr and $t = 1.014$~Myr, respectively. 
The color shows the magnitude of velocity in a given cell. Different symbols 
denote the position of the cell: $r < R_\mathrm{SC}$ (x), 
and $r > R_\mathrm{SC}$ (dot). Note that the instabilities develop within 
$R_\mathrm{st} = 8.991$pc, $R_\mathrm{st} = 9.696$~pc and $R_\mathrm{st}
= 1.860$~pc in the case of models 5, 6 and 8, respectively. Dashed lines 
enclose the region from which the line profiles are computed. Right:
Corresponding line profiles coming from $r < R_\mathrm{SC}$ (red), $r >
R_\mathrm{SC}$ (green) and their sum (blue).
}
\label{m5_lines}
\end{figure}


\begin{thebibliography}{99}

\bibitem{1} Beck, S. C., Turner, J. L., Langland-Shula, L. E., Meier, D. S.,
            Crosthwaite, L. P. \& Gorjian, V. 2002, AJ, 124, 2516
\bibitem{2} Beck, S. C. 2008, A\&A, 489, 567

\bibitem{3} Burke, J. R. \& Silk, J. 1974, ApJ, 190,1

\bibitem{4} Cant{\'o}, J., Raga, A. C. \&  Rodriguez, L. F. 2000, ApJ, 536,
            896
\bibitem{5} Chevalier, R. A. \& Clegg, A. W. 1985, Nature, 317, 44 
\bibitem{6} De Pree, C. G., Wilner, D. J., Goss, W. M., Welch, W. J.
            \& McGrath, E. 2000, ApJ, 540, 308
\bibitem{7} Galliano, E. \& Alloin, D. 2008, A\&A, 487, 519
\bibitem{8} Gilbert, A. M., Graham, J. R., McLean, I. S. et al. 2000,
            ApJ, 533, L57
\bibitem{9} Gilbert, A. M. \& Graham, J. R. 2007, ApJ, 668, 168
\bibitem{10} Henry, A. L., Turner, J. L., Beck, S. C., Crosthwaite, L. P. \& 
            Meier, D. S. 2007, AJ, 133, 757
\bibitem{11} Leitherer, C., Schaerer, D., Goldader, J.D. et al., 
            1999, ApJS, 123, 3
\bibitem{12} Murray, N., Quataert, E. \& Thompson, T. A. 2005, ApJ, 618, 569 
\bibitem{13}  Palou\v{s}, J., W\"unsch, R., Tenorio-Tagle, G. \& Silich, S.
             2008, IAUS, 254, 233
\bibitem{14} Rodr\'\i guez-Rico, C. A., Goss, W. M., Turner, J. L. \& 
             G\'omez, Y. 2007, ApJ, 670, 295
\bibitem{15} Shapiro, P. R. \& Kang, H. 1987, ApJ, 318, 32
\bibitem{16} Silich, S., Tenorio-Tagle, G. \& Mu\~noz-Tu\~n\'on, C. 2007,
             ApJ, 669, 952
\bibitem{17} Silich, S., Tenorio-Tagle G. \& Rodr\'\i{}guez Gonz\'alez, A. 
             2004, ApJ, 610, 226 
\bibitem{18} Silich, S., Tenorio-Tagle, G., Torres-Campos, A., 
             Mu\~noz-Tu\~n\'on, C., Monreal-Ibero, A. \& Melo, V. 2009,
             ApJ, 700, 931  
\bibitem{19} Silk, J. \& Burke, J. R. 1974, ApJ, 190, 11
\bibitem{20} Tenorio-Tagle, G. 1996, AJ, 111, 1641
\bibitem{21} Tenorio-Tagle, G., Silich, S., Rodr\'iguez-Gonz\'alez A. 
             \& Mu\~noz-Tu\~non, C., 2005, ApJ Lett. 628, L13
\bibitem{22} Tenorio-Tagle, G., W\"unsch, R., Silich, S. \& Palou\v{s}, J.
             2007, ApJ, 658, 1196
\bibitem{23} Turner, J.L., Beck, S.C. \& Ho, P.T.P. 2000, ApJ, 532,
             L109
\bibitem{24} Turner, J. L., Beck, S. C., Crosthwaite, L. P. et al.
             2003, Nature, 423, 621
\bibitem{25} Vietri, M. \& Pesce, E. 1995, ApJ, 442, 618
\bibitem{26} W\"unsch, R.,  Silich, S. Palou\v{s}, J. \& Tenorio-Tagle, G.
             2007, A\&A, 471, 579
\bibitem{27} W\"unsch, R.,  Tenorio-Tagle, G., Palou\v{s}, J. \& Silich, S.
             2008, ApJ, 683, 683
\bibitem{28} Zeldovich, Ya. B. \& Raizer, Yu. P. Physics of Shock Waves and 
             High-Temperature Hydrodynamic Phenomena, vol. 2 NY: Academic

\end{thebibliography}
\end{document}